\documentclass[%
 reprint,
superscriptaddress,
 amsmath,amssymb,
 aps,twocolumn,
pra,floatfix
]{revtex4-1}

\usepackage{graphicx}
\usepackage{dcolumn}
\usepackage{bm}
\usepackage{dsfont}
\usepackage{blindtext}
\usepackage{mathtools}

\newcommand\varpm{\mathbin{\vcenter{\hbox{%
  \oalign{\hfil$\scriptstyle+$\hfil\cr
          \noalign{\kern-.3ex}
          $\scriptscriptstyle({-})$\cr}%
}}}}

\begin{document}


\title{Interferometric Approach to Open Quantum Systems and Non-Markovian Dynamics}

\author{Olli Siltanen}
\affiliation{Turku Centre for Quantum Physics, Department of Physics and Astronomy, University of Turku, FI-20014 Turun yliopisto, Finland}
\affiliation{QTF Centre of Excellence, Department of Physics and Astronomy, University of Turku, FI-20014 Turun yliopisto, Finland}
\affiliation{Laboratory of Quantum Optics, Department of Physics and Astronomy, University of Turku, FI-20014 Turun yliopisto, Finland}

\author{Tom Kuusela}
\affiliation{Turku Centre for Quantum Physics, Department of Physics and Astronomy, University of Turku, FI-20014 Turun yliopisto, Finland}
\affiliation{QTF Centre of Excellence, Department of Physics and Astronomy, University of Turku, FI-20014 Turun yliopisto, Finland}
\affiliation{Laboratory of Quantum Optics, Department of Physics and Astronomy, University of Turku, FI-20014 Turun yliopisto, Finland}

\author{Jyrki Piilo}
\affiliation{Turku Centre for Quantum Physics, Department of Physics and Astronomy, University of Turku, FI-20014 Turun yliopisto, Finland}
\affiliation{QTF Centre of Excellence, Department of Physics and Astronomy, University of Turku, FI-20014 Turun yliopisto, Finland}
\affiliation{Laboratory of Quantum Optics, Department of Physics and Astronomy, University of Turku, FI-20014 Turun yliopisto, Finland}



\date{\today}

\begin{abstract}

We combine the dynamics of open quantum systems with interferometry and interference  introducing the concept of open system interferometer. By considering a single photon in a Mach-Zehnder interferometer, where the polarization (open system) and frequency (environment) of the photon interact, we theoretically show how inside the interferometer path-wise polarization dephasing dynamics is Markovian while the joint dynamics displays non-Markovian features. Outside the interferometer and due to interference, the open system displays rich dynamical features with distinct alternatives: Only one path displaying non-Markovian memory effects, both paths individually displaying them, or no memory effects appearing at all. The scheme allows (1) to probe the optical path difference inside the interferometer by studying path-wise non-Markovianity outside the interferometer, and (2) to introduce path-wise dissipative features for the open system dynamics even though the system-environment interaction itself contains only dephasing. Due to the path-dependencies, our results are tightly connected to quantum erasure. In general, our results open so far unexplored avenues to control open system dynamics and for fundamental studies of quantum physics.

\end{abstract}

\maketitle



\section{Introduction}
Interactions within a multipartite quantum system can destroy the quantum properties of a given subsystem. This leads to disturbance, or decoherence, in the system of interest, i.e., an open quantum system~\cite{petru}. The study of decoherence and open quantum systems in general is important for both practical and fundamental reasons, e.g., to produce feasible quantum devices harnessing the fragile properties threatened by the environment~\cite{suter}, or to better understand such essentials as quantum to classical transition~\cite{zurek,blanchard,mazzola}  and non-Markovian character of open system evolution~\cite{rivas,colloquium,devega,li,meaning,good_for,NM_to_M_1,NM_to_M_2,BLP,NM_to_M_3,M_to_NM,achiu,fff,Ber2015,scia1,sau}.

Linear optical systems provide a commonly used practical platform for this open system framework. Here, the system of interest is often the polarization of photon while the environment is the frequency degree of freedom. The system-environment interaction is due to a birefringent medium and the subsequent polarization-frequency coupling~\cite{M_to_NM,synthetic,probing,telep,tele2,SDC,sina,partitions,nonlocal_memory,photonic_real}. 
Recent achievements within this framework include, e.g.,  controlled Markovian to non-Markovian transition~\cite{M_to_NM} and arbitrary control of the dephasing dynamics~\cite{synthetic}. Sometimes the frequency noise can even turn out to be useful. For example, it has been shown that noise-induced non-Markovianity can be exploited in teleportation~\cite{telep,tele2} and  superdense coding~\cite{SDC}.

Our current aim is to go beyond the conventional open quantum system framework by combining the system-environment interaction scheme with interferometric studies of quantum optics, i.e., to introduce the concept of {\it{open system interferometer}}.  
We are interested in how noise appearing in different locations of the interferometer influences its output. At the same time we describe how the interferometric setup influences the dynamics of open quantum system and the appearance of non-Markovian memory effects. These have been under intensive scrutiny both theoretically and experimentally in the last ten years~\cite{rivas,colloquium,devega,li,meaning,good_for}, though not yet considered in the interferometric framework to the best of our knowledge.

In addition to the frequency of the photon,  the paths of the interferometer introduce another environmental degree of freedom and allow to apply 
noise in different locations of the interferometer---both inside and outside.  By considering a Mach-Zehnder interferometer, we see how prior noise influences the interference at the output and the subsequent open system dynamics. In general and as a result of this ``interferometric reservoir engineering", we obtain non-Markovian memory-effects depending on in which location of the interferometer the state tomography is performed and where the ``Heisenberg cut"~\cite{heisenberg_cut}  between the system and the environment is drawn. The framework, due to interference, also allows to mimic dissipative features of open system dynamics. While previous work has utilized wave plates~\cite{salles}, we achieve dissipative-like dynamics less trivially, since we are dealing with pure dephasing that leaves the polarization probabilities invariant. Moreover, our model can be seen as an extension of optical collision models. Typically, the dynamics between the collisions (described in the linear optical framework by beam splitters) is unitary and discrete~\cite{cm1,cm2,cm3}. Here, we account for non-unitary and continuous-time dynamics.

It is worth noting that earlier works have studied the problematics of convex combinations of dynamical maps (see, e.g., Refs.~\cite{sau,vina}) and  how combining quantum channels in different causal orders allows to improve information transmission for communication purposes (see, e.g.,  Refs.~\cite{co1,co2}). However, our motivation and interest are different. We are interested in the fundamental studies of open quantum systems and non-Markovian features, when combining dynamical maps coherently in an interferometric setup. That is, when the channels are temporally aligned and the path difference concerning free evolution is zero. This also allows us to access the interferometric effects that are often ignored in incoherent mixing~\cite{sau,salles}.

\begin{figure}[t]
\centering
\includegraphics[width=1\linewidth]{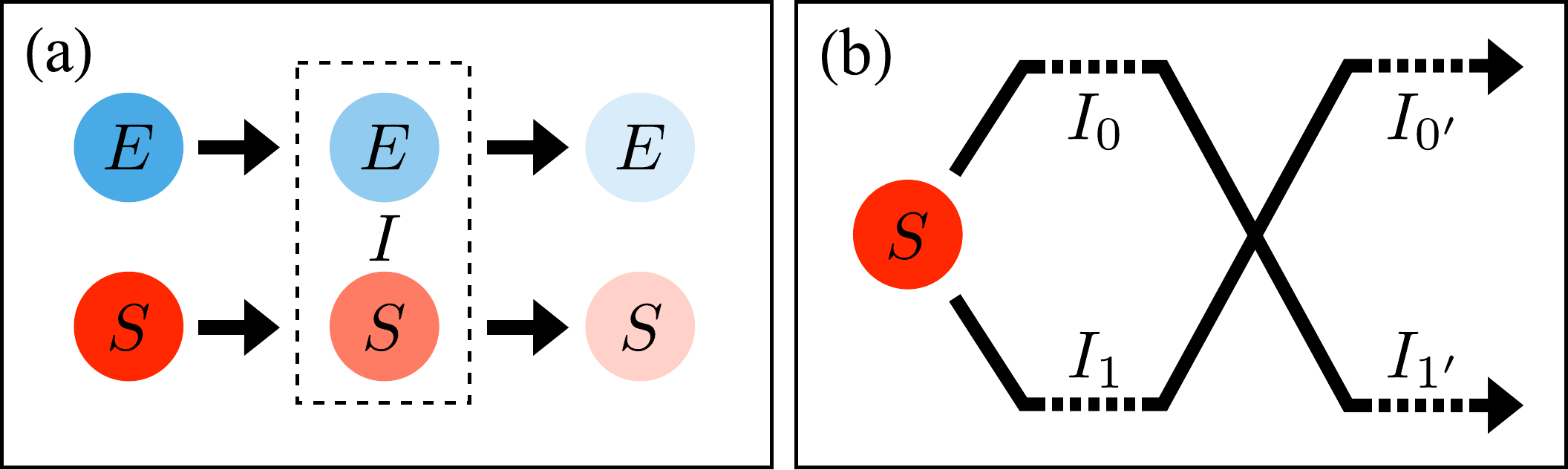}
\caption{(Color online) A schematic picture of the open system (polarization) and environment (frequency + path) studied in this paper. Panel (a) depicts the relationship between the system $S$ and environment $E$ in the standard linear optical approach, i.e., when only frequency is included in the environment, whilst the path degree of freedom and its relationship with polarization and frequency is illustrated in panel (b). $I$ stands for interaction.}
\label{model}
\end{figure}

We consider a polarization qubit of a single photon experiencing frequency noise on the two paths in and outside a Mach-Zehnder interferometer.
A schematic picture of this model is presented in Fig.~\ref{model}.  In panel (a), we have the conventional open system view with unitary coupling between the polarization and frequency causing dephasing. In panel (b), the unitaries with possibly different interaction times and refractive indices  are applied on the different paths of the interferometer. Throughout this paper, we use labels 0 and 1 for the paths inside the interferometer, and $0^{'}$ and $1^{'}$ for the paths outside the interferometer.

Intuitively, same unitaries on paths $0^{(')}$ and $1^{(')}$ should not alter the open system dynamics from the traditional single-path case, nor should they affect interference, since interference is strongly related to the indistinguishability of the paths. Thus, the two main questions of this paper are: \textit{How do different path-wise unitaries affect the total dynamics and interference? How does interference, in turn, affect the following dynamics?} We will address these questions from the point of view of both the total open system state and the conditional path-wise states, revealing the intriguing effects related to the quantum erasure and the which-path-information, respectively. Next, we briefly recall the system-environment interaction model.

\section{System-environment interaction and information flow}
Omitting the path qubit for now, the initial polarization-frequency state is
\begin{equation}
|\Psi\rangle=C_H|H\rangle\int d\omega g(\omega)e^{i\theta_H}|\omega\rangle+C_V|V\rangle\int d\omega g(\omega)e^{i\theta_V}|\omega\rangle,
\label{initial_state}
\end{equation}
where $C_{H(V)}$ and $g(\omega)$ are the probability amplitudes for the photon to be in the polarization state $|H(V)\rangle$ and the frequency state $|\omega\rangle$, respectively, and $e^{i\theta_{H(V)}}$ is the complex phase factor corresponding to horizontal (vertical) polarization. Note that in general $\theta_{H(V)}=\theta_{H(V)}(\omega)\neq$ constant, indicating initial correlations between the polarization and frequency~\cite{synthetic,sina,partitions}. Here, however, we restrict ourselves to constant initial phase factors and initial product state between the system and environment.

Individually, the action of the dephasing channels on the system is well-known~\cite{M_to_NM,synthetic,probing,telep,SDC,sina,partitions,colloquium,good_for,
nonlocal_memory,photonic_real}. In the linear optical framework, we have
$\varrho_j(t)=\Phi_j(t)\big(\varrho(0)\big)
=\text{tr}_E[U_j(t)|\Psi\rangle\langle\Psi|U_j(t)^{\dagger}]$,
where
$U_j(t)|\lambda\rangle|\omega\rangle=e^{in_{j\lambda}\omega T_j(t)}|\lambda\rangle|\omega\rangle$
and $\lambda$ labels the polarization components $\{H,V\}$ while $j$ is the channel label.
In the time evolution operator $U_j$ we have
$T_j(t)=\int_0^t\chi_j(s)ds$ with
$\chi_j(s)=1$, when $t_{ji}\leq s \leq t_{jf}$, and $\chi_j(s)=0$ otherwise.
Thereby, the polarization and frequency are coupled in a birefringent medium described by the refractive indices $n_{j\lambda}$ from time $t_{ji}$ to $t_{jf}$.

Employing a Gaussian frequency distribution~\cite{M_to_NM}
$|g(\omega)|^2=\frac{1}{\sqrt{2\pi\sigma^2}}\text{exp}\Big[-\frac{1}{2}\Big(\frac{\omega-\mu}{\sigma}\Big)^2\Big],$
the evolving state of the open system is
\begin{equation}
\varrho_j(t)=\renewcommand{\arraystretch}{1.3}
\begin{pmatrix}
|C_H|^2 & C_HC_V^*\kappa_j(t) \\
C_H^*C_V\kappa_j(t)^* & |C_V|^2
\end{pmatrix},
\label{density1}
\end{equation}
where the coherence terms undergo rotation and decay dictated by the decoherence function
\begin{equation}
\kappa_j(t)=\text{exp}\Big[i\big(\theta+\mu\Delta n_jT_j(t)\big)-\frac{1}{2}\big(\sigma\Delta n_jT_j(t)\big)^2\Big].
\label{decoherence}
\end{equation}
Here, $\Delta n_j=n_{jH}-n_{jV}$ is the birefringence of the medium, and $\theta=\theta_H-\theta_V$.

The flow of information between the system and environment---and its connection to non-Markovian dynamics---is commonly described by the trace distance $D(t)$ between a pair of initially distinguishable states of the system~\cite{BLP} and has been applied in several physical contexts for this purpose in the past, see, e.g., Refs.~\cite{wu,smirne,fff,wissmann,dong,yu,haase}. The sign of $\frac{d}{dt}D(t)$ tells the direction of the information flow.  Positive sign
indicates non-Markovian memory effects and information backflow into the open system. For dephasing and choosing the initial state pair to be $|\pm\rangle=\frac{1}{\sqrt{2}}(|H\rangle\pm|V\rangle)$, i.e., having the maximum initial coherences, the trace distance has a simple expression $D_j(t)=|\kappa_j(t)|$~\cite{M_to_NM}. Hence, in our case, information backflow manifests itself by increasing coherences.

It should be stressed that there are many more indicators of non-Markovianity~\cite{petru,rivas,colloquium,devega,li,meaning,good_for,teittinen,ocp,budini1,budini2,budini3,budini4}. Most notably, the monotonicity of the trace distance coincides with CP-divisibility in single qubit dephasing, as we will show later. Some recently proposed definitions of quantum Markovianity are related to stochastic quantum processes including intermediate control operations and measurements~\cite{ocp,budini1,budini2,budini3,budini4}. Despite our system of interest not falling into this category, CP-divisibility coincides with the stricter, \textit{operational divisibility} on average~\cite{sau,ocp}, and so does the trace distance too, in our case.

\section{The interferometric setup}
A balanced convex combination of two of the channels  can be constructed, e.g., by a Mach-Zehnder type interferometer, inside of which the channels $\Phi_0(t)$ and $\Phi_1(t)$ operate on their own paths. Including the path of the photon---initially in the state $|\tilde{0}\rangle$, not to be confused with the path states $|0\rangle$ and $|0'\rangle$---in the environment, the overall polarization-frequency-path state inside the interferometer is
\begin{equation}
\begin{split}
|MZ(t)\rangle&:=(U_0(t)\otimes|0\rangle\langle0|+U_1(t)\otimes|1\rangle\langle1|)(\mathds{1}\otimes H)|\Psi\rangle|\tilde{0}\rangle\\
&=\frac{1}{\sqrt{2}}(U_0(t)|\Psi\rangle|0\rangle+U_1(t)|\Psi\rangle|1\rangle),
\label{MZ}
\end{split}
\end{equation}
where, to see how the system-environment interaction affects interference, we
have assumed that there is no phase difference between the paths, and $H$ is the Hadamard gate describing a non-polarizing 50/50 beam splitter. From Eq.~\eqref{MZ} it is clear that obtaining the which-path-information, i.e., applying $\mathds{1}\otimes|j\rangle\langle j|$ and normalizing the state, results in Markovian dephasing dynamics of the system when Gaussian frequency distribution is used. However, as long as the path is \textit{not} measured, we can go beyond Markovian dynamics. The state of the system in the latter case is given by
\begin{equation}
\varrho(t)=\frac{\Phi_0(t)\big(\varrho(0)\big)+\Phi_1(t)\big(\varrho(0)\big)}{2}.
\label{no_measurement}
\end{equation}

\begin{figure}[t]
\centering
\includegraphics[width=\linewidth]{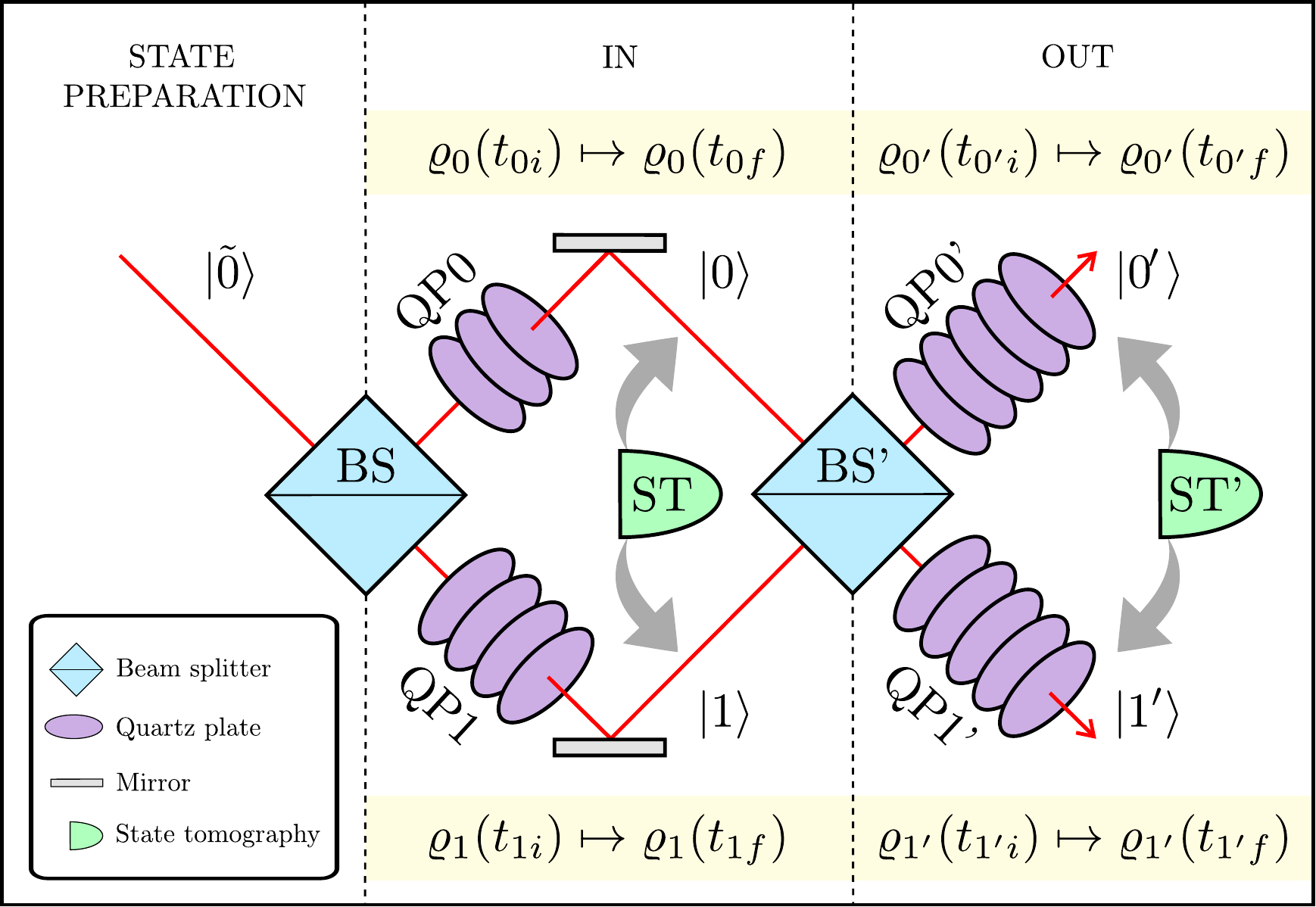}
\caption{(Color online) A setup realizing the open system dynamics described by $\varrho_j(t_{ji})\mapsto\varrho_j(t_{jf})$ and $\varrho_{j'}(t_{j'i})\mapsto\varrho_{j'}(t_{j'f})$. The interaction times are controlled by varying the thicknesses of the corresponding quartz plates. To obtain the path-wise dynamics we perform state tomography on the desired path. Total dynamics is the sum of these transformations weighted by the path probabilities.}
\label{interferometer}
\end{figure}

Now the question becomes, what kind of open system dynamics 
we have  \textit{after} the interferometer, both on the individual paths separately and combining them. This time, for simplicity, we have the same unitary coupling $U'(t):=U_{0'}(t)=U_{1'}(t)$ acting after both exit ports. The total state exiting the interferometer is
\begin{equation}
\begin{split}
|MZ'(t)\rangle:&=(U'(t)\otimes\mathds{1})(\mathds{1}\otimes H)|MZ(t)\rangle\\
&=\frac{1}{2}\big[U'(t)\big(U_0(t)+U_1(t)\big)|\Psi\rangle|0'\rangle\\
&\hspace{13pt}+U'(t)\big(U_0(t)-U_1(t)\big)|\Psi\rangle|1'\rangle\big].
\label{MZ'}
\end{split}
\end{equation}
The open system state $\varrho'(t)$ is then given by Eq.~\eqref{no_measurement} with the transformation
\begin{equation}
n_{j\lambda}T_j(t)\mapsto n_{j\lambda}T_j(t)+n_\lambda'T'(t)
\label{last_unitary_coupling}
\end{equation}
applied to it. However, if we \textit{now} measure the photon's path and obtain the result $j'$, the state of the system becomes
\begin{equation}
\renewcommand{\arraystretch}{1.3}
\varrho_{j'}(t)=\frac{1}{4P_{j'}}\Bigg[2\varrho'(t)+(-1)^{j'}
\begin{pmatrix}
|C_H|^2\kappa_H & C_HC_V^*\Lambda(t)\\
C_H^*C_V\Lambda(t)^* & |C_V|^2\kappa_V
\end{pmatrix}\Bigg],
\label{density2}
\end{equation}
where
\begin{equation}
\kappa_\lambda=2\hspace{3pt}\text{exp}\Big[-\frac{1}{2}\sigma^2(n_{0\lambda}t_0-n_{1\lambda}t_1)^2\Big]\cos\big[\mu(n_{0\lambda}t_0-n_{1\lambda}t_1)\big]
\label{interference_kappa}
\end{equation}
and
\begin{equation}
\begin{split}
\Lambda(t)=&\text{exp}\Big\{i\big[\theta+\mu\big(n_{0H}t_0-n_{1V}t_1+\Delta n'T'(t)\big)\big]\\
&\hspace{20pt}-\frac{1}{2}\sigma^2\big(n_{0H}t_0-n_{1V}t_1+\Delta n'T'(t)\big)^2\Big\}\\
+&\text{exp}\Big\{i\big[\theta+\mu\big(n_{1H}t_1-n_{0V}t_0+\Delta n'T'(t)\big)\big]\\
&\hspace{20pt}-\frac{1}{2}\sigma^2\big(n_{1H}t_1-n_{0V}t_0+\Delta n'T'(t)\big)^2\Big\}
\end{split}
\label{interference_lambda}
\end{equation}
originate from the cross-terms $U_0(t)|\Psi\rangle\langle\Psi|U_1(t)^\dagger$
and  $U_1(t)|\Psi\rangle\langle\Psi|U_0(t)^\dagger$ while
\begin{equation}
P_{j'}=\frac{2+(-1)^{j'}|C_H|^2\kappa_H+(-1)^{j'}|C_V|^2\kappa_V}{4}
\label{new_prob_j}
\end{equation}
is the probability for the photon to be detected on path $j'$ outside the interferometer. 
Note that both the polarization probabilities in the path-wise states  [Eq.~(\ref{density2})] and the path probabilities $P_{j'}$ [Eq.~(\ref{new_prob_j})] contain rapidly
oscillating terms with the frequency $\mu$, since they both contain $\kappa_{\lambda}$ [Eq.~(\ref{interference_kappa})]. 
An experimental setup which can be used to realize both the path-wise and joint open system dynamics is presented in Fig.~\ref{interferometer}.

\section{Dynamical characteristics of the open system interferometer}

Here, we study the distinguishability of states in more detail. As mentioned earlier, trace distance coincides with CP-divisibility as an indicator of non-Markovian memory effects  in the case of a single qubit dephasing channel, whenever a completely positive and trace preserving (CPTP) map exists. This is trivially the case on paths 0 and 1 and when averaged over paths either before or after the second beam splitter, BS'. However, the existence of a CPTP map is a more subtle issue with the conditional state $\varrho_{j'}(t)$ on path $j'$ described by Eq.~\eqref{density2}. Interpreting $\varrho_{j'}(0)$ as the initial state instead of $\varrho(0)$---and thus interpreting the preceding interaction times $t_0$ and $t_1$ as fixed parameters related to state preparation---yields a CPTP map for some parameter choices but not for all due to the contractivity requirement of the trace distance, $D(t)\leq D(0)$. On the other hand, the $\varrho(0)$-dependent normalization constant $P_{j'}$ prevents one of having a CPTP map with the initial state $\varrho(0)$ either. Still, we argue that the trace distance is a valid indicator of non-Markovianity even on paths 0' and 1' individually; Consider a non-normalized version of Eq.~\eqref{density2}. This is a completely positive and trace non-increasing (CPTNI) \textit{quantum operation} $\mathcal{E}_{j'}(t)$ \cite{chuang} having the Kraus representation $\mathcal{E}_{j'}(t)\big(\varrho(0)\big)=\sum_{i=0}^1K_{i,j'}(t)\varrho(0)K_{i,j'}(t)^\dagger$ with the Kraus operators
\begin{equation}
\sqrt{\frac{\sqrt{h_{j'}v_{j'}}\pm|f_{j'}(t)|}{2\sqrt{h_{j'}v_{j'}}}}
\begin{pmatrix}
\pm\sqrt{h_{j'}}\frac{f_{j'}(t)}{|f_{j'}(t)|}&0\\
0&\sqrt{v_{j'}}\end{pmatrix},
\label{Kraus_i}
\end{equation}
where $h_{j'}=[2+(-1)^{j'}\kappa_H]/4$, $v_{j'}=[2+(-1)^{j'}\kappa_V]/4$, $f_{j'}(t)=[\kappa_0(t)+\kappa_1(t)+(-1)^{j'}\Lambda(t)]/4$, and $\sum_{i=0}^1K_{i,j'}(t)^\dagger K_{i,j'}(t)\leq\mathds{1}$. $\sum_{i=0}^1K_{i,j'}(t)^\dagger K_{i,j'}(t)=\mathds{1}$ holds for $h_{j'}=v_{j'}=1$, making $\mathcal{E}_{j'}(t)$ a valid CPTP map.

Quantum dynamics is often considered Markovian if it can be split according to $\Phi(t_2)=V(t_2,t_1)\Phi(t_1)$, where $t_2\geq t_1\geq 0$ and both $\Phi(t)$ and the intermediate propagator $V(t_2,t_1)$ are CPTP \cite{petru}. Let us instead consider the criterion's generalized version, where the operators are CPTNI. In our case, the path-wise propagator $V_{j'}(t_2,t_1)$ can be defined via its (two) Kraus operators
\begin{equation}
\sqrt{\frac{1\pm\frac{|f_{j'}(t_2)|}{|f_{j'}(t_1)|}}{2}}
\renewcommand*{\arraystretch}{1.5}\raisebox{0.5em}{$\begin{pmatrix}
\pm\frac{f_{j'}(t_2)}{f_{j'}(t_1)}\frac{|f_{j'}(t_1)|}{|f_{j'}(t_2)|}&0\\
0&1\end{pmatrix}$}.
\label{Kraus_2}
\end{equation}
$V_{j'}(t_2,t_1)$ is not only CPTNI but CPTP if and only if $|f_{j'}(t_2)|\leq|f_{j'}(t_1)|$, which is equivalent with $\frac{d}{dt}|f_{j'}(t)|\leq0$. Therefore, trace distance coincides with CP-divisibility and is a valid indicator of non-Markovianity (in the generalized CPTNI sense) also on paths 0' and 1'. The trace decreasing part of the dynamics occurs strictly at BS' and hence needs not be considered in $V_{j'}(t_2,t_1)$. In fact, $V_{j'}(t_2,t_1)$ is also the propagator in the traditional CPTP case, as it is independent of the parameters $h_{j'}$ and $v_{j'}$. It should be stressed that, in general, there is no clear definition of quantum Markovianity regarding non-CP or non-TP maps.

\begin{figure}[t!]
\centering
\includegraphics[width=\linewidth]{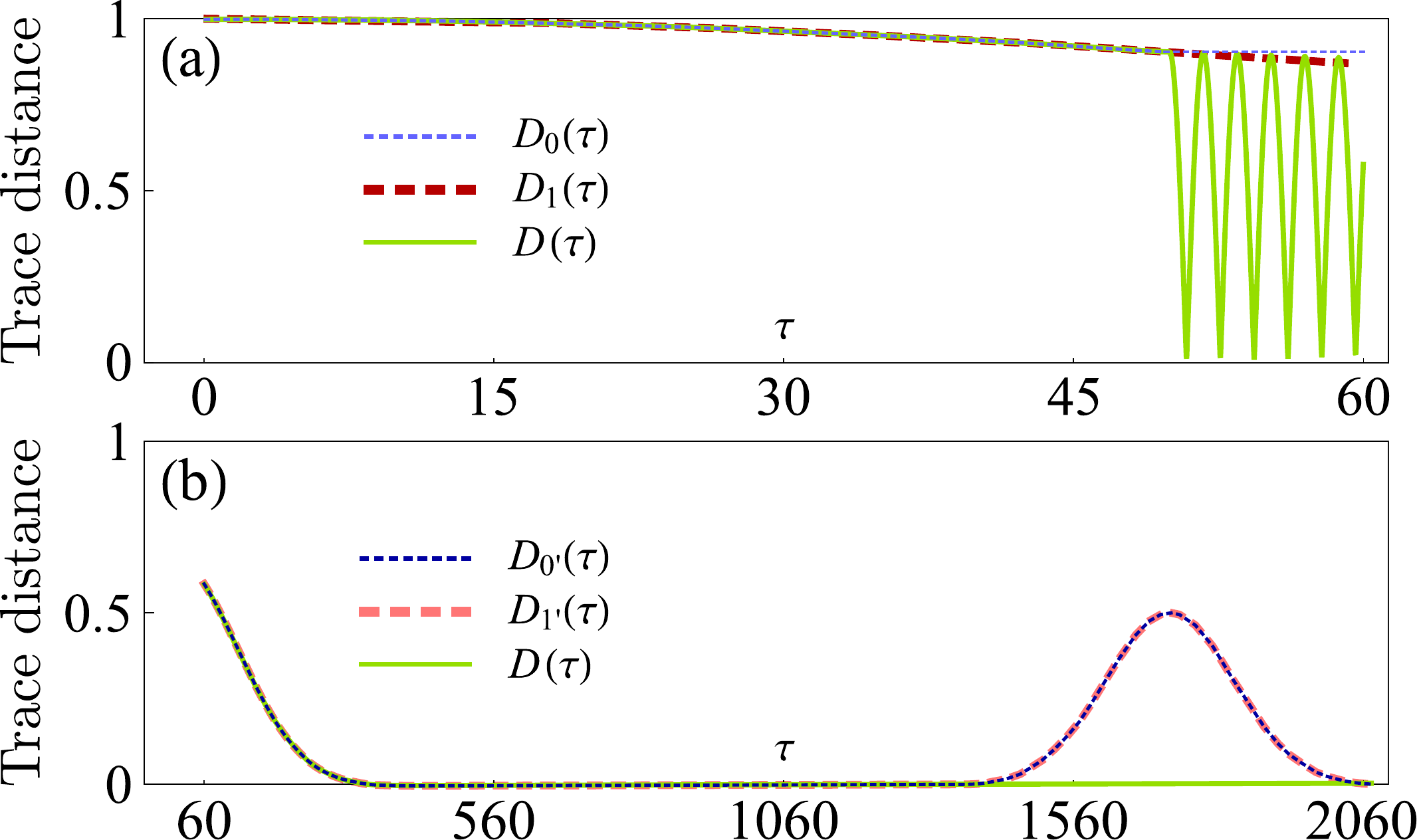}
\caption{(Color online) Trace distances of the initial state pair $|\pm\rangle$ (a) in and (b) outside the interferometer as functions of the scaled laboratory time $\tau$ when  $|\Delta \tau| = 10$. Dashed light blue = path 0; dashed and thick dark red = path 1; dashed dark blue = path 0'; dashed and thick light red = path 1'; solid green = combined paths dynamics. We have fixed $n_H=1.553$, $n_V=1.544$, $\mu/\sigma=400$, $\tau_0=50$, and  $\tau_1=60$. For the dynamics outside the interferometer in panel (b), the interaction times on both output paths start simultaneously at $\tau=60$ and then run freely.}
\label{results}
\end{figure}

We consider the case where the polarization-specific refractive indices are the same but  the interaction times inside the interferometer may differ. Interaction times on the paths outside the interferometer are equal.
In terms of notation, $t$ is the laboratory time, $t_0=t_{0f}-t_{0i}$ ($t_1=t_{1f}-t_{1i}$) is the duration of the interaction on path $0$ ($1$),
and we use $t_{0i}=t_{1i}=0$. The difference in the interaction times inside the interferometer is denoted with $\Delta t = t_0-t_1$. For time scales we use $\tau=\sigma t$ and in similar manner have $\tau_0=\sigma t_0,~\tau_1=\sigma t_1$, and $\Delta \tau = \sigma \Delta t$.  

We first consider  the case where the interaction time difference, $|\Delta \tau|=10$, is so large that the subsequent optical path differences produced inside the interferometer prevent interference at BS'. We have plotted the trace distances capturing the non-Markovian features of the dephasing dynamics in Fig.~\ref{results}, the initial state pair being $|\pm\rangle$. Taking both paths into consideration (i.e., implementing ``quantum erasure") yields $D(\tau)$, whereas performing state tomography only on path $j^{(')}$ yields $D_{j^{(')}}(\tau)$.  Figure~\ref{results} (a) shows that inside the interferometer before BS', the joint open system undergoes non-Markovian dephasing, while the path-wise states behave in a Markovian fashion. As soon as the interaction on path 0 is switched off at $\tau=50$---while interaction still continues on path 1---the joint open system dynamics displays oscillatory behaviour of trace distance indicating non-Markovian behaviour.  Outside the interferometer, see Fig.~\ref{results} (b),  the open system displays opposite features. Here, the joint dynamics is Markovian while the path-wise evolution shows non-Markovianity and backflow of information. Initially on each output path, there are $H$ and $V$ components originating from both paths inside the interferometer. In the subsequent interaction outside the interferometer and on each path, the optical path differences between the $H$ component from one earlier path and the $V$ component from the other earlier path become temporarily equal allowing recoherence and memory effects to arise. The maximum trace distance reached is equal to $0.5$ since the other two remaining  components have distinct path differences at all times. 

It is also interesting to note that $\Lambda(t)$, which gives the path-wise dynamics outside the interferometer,
contains information about the system's entire history---see Eqs.~(\ref{density2}) and (\ref{interference_lambda}). 
This observation leads to the following result. When there is no interference at BS', 
we can estimate that
\begin{equation}
|t_0-t_1|\approx\frac{|\Delta n|t_\text{max}}{\max\{n_H,n_V\}},
\label{time_diff}
\end{equation}
where $t_\text{max}$ is the instant of total interaction time where $|\Lambda(t)|$ reaches its (observable) maximum. Therefore, by studying non-Markovianity outside the interferometer, we can quantitatively estimate what the interaction time difference was inside the interferometer---even though the path probabilities $P_{0'}$ and $P_{1'}$
do not carry significant information about this anymore. 
If interaction times are equal along the two paths and instead indices of refraction are not equal, 
we can estimate their difference in the same way.
Similar calculations also hold for estimating $|C_H|$, $|C_V|$, and their relative phase $\theta$. Note that, with certain parameters, the path-wise coherences outside the interferometer start from very close to zero. In these scenarios, analyzing non-Markovianity is not just an alternative way but the only way to make these estimations.

\begin{figure}[t!]
\centering
\includegraphics[width=0.9\linewidth]{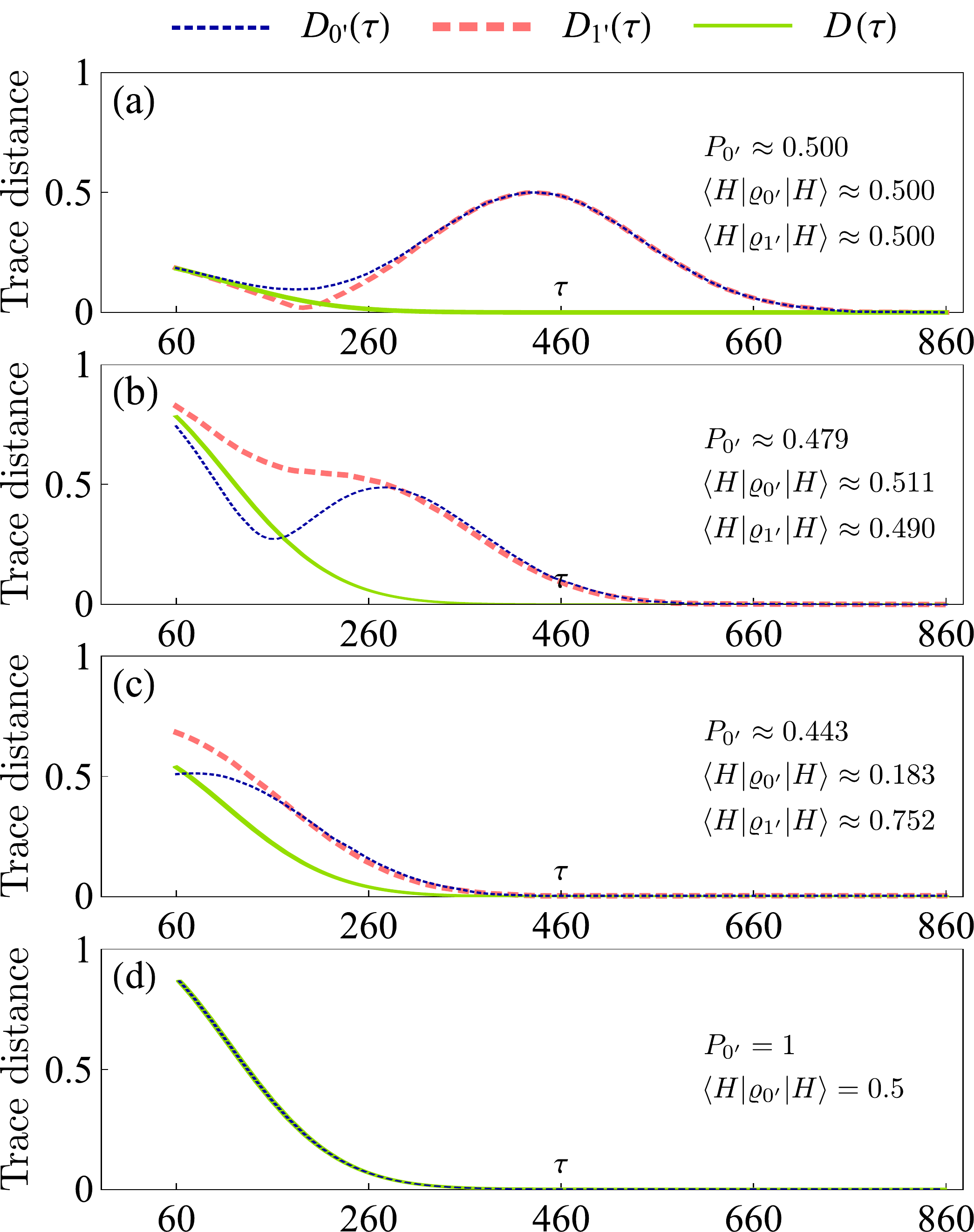}
\caption{(Color online) Trace distance dynamics outside the interferometer for different values of $|\Delta \tau|$.
(a) $|\Delta \tau| = 2.5 $, (b) $|\Delta \tau| = 1.5$, (c) $|\Delta \tau| = 0.5$, and (d) $|\Delta \tau| = 0 $.
 Other parameters, notation, and units are the same as in Fig.~\ref{results} (b), except:
 (a) $\tau_0=57.5$,  (b) $\tau_0=58.5$,  (c) $\tau_0=59.5$, and  (d) $\tau_0=60$.}
 
\label{small_diff}
\end{figure}

Let us now turn to the question on how the increasing amount of interference at BS' of the interferometer influences 
the subsequent open system dynamics in the joint and path-wise states. The results are shown in Fig.~\ref{small_diff},
where from panel (a) to (d) we have $|\Delta \tau| = 2.5$ to  $|\Delta \tau| = 0$, respectively.
Comparing Fig.~\ref{results} (b) having no interference ($|\Delta \tau| = 10$) and Fig.~\ref{small_diff} (a) ($|\Delta \tau| = 2.5$) we see that the recoherence peak and the interval of non-Markovianity shift to smaller times $\tau$  and that the behaviour of the path-wise state dynamics begin to deviate from themselves even though both still display memory effects. 
Increasing the amount of interference further and having $|\Delta \tau|=1.5$ in Fig.~\ref{small_diff} (b) shows that the dynamics on path 0' displays information backflow while on path 1' and joint dynamics behave in Markovian manner. 
Note also that in the path-wise states, the probabilities $\langle H|\varrho_{0'} | H \rangle$ and  $\langle H|\varrho_{1'} | H \rangle$ have changed compared to their initial value $0.5$. This means that the interferometric setup also allows to introduce dissipative-type effects for the open system dynamics, due to interference, even though the system-environment interaction consists of only dephasing. This is seen in more significant way in  Fig.~\ref{small_diff} (c) with $|\Delta \tau|=0.5$. Here, we have, e.g., on path 0' $\langle H|\varrho_{0'} | H \rangle\approx 0.183$  and at the same time all the three different dynamics behave in Markovian way, though distinctively. Finally,  Fig.~\ref{small_diff} (d)
($|\Delta \tau|=0$) represents the other extreme compared to Fig.~\ref{results} (b). Here, despite of having noise inside the interferometer, the two previous paths are fully indistinguishable, and due to full interference, the photon always ends up to path 0' and no memory effects are on display.


\section{Conclusions}

We have gone beyond the traditional view point of open quantum system dynamics by introducing and studying open system interferometer. By considering a single photon in a Mach-Zehnder interferometer and accounting for polarization-frequency interaction at different stages of the interferometer, we have shown how, inside the interferometer, the path-wise dephasing dynamics of the open system (polarization) displays Markovian dynamics, while the joint dynamics including both of the paths displays non-Markovian memory effects---a direct result of quantum erasure, i.e., ignoring the path. More importantly and interestingly, at the output of the interferometer, we observe a subtle and rich interplay between the interference and memory effects. Depending on the system-environment interaction times inside the interferometer, the open system dynamics in the output can display non-Markovianity and information backflow only on one path, on both paths individually, or no memory effects at all. At this point, quantum erasure concerns not only which-path-information but also memory effects, given that the unitaries are equal. It is also important to note that the scheme can be used to estimate the optical path difference inside the interferometer by looking at non-Markovianity at the output---while the path probabilites do not carry this information anymore. Moreover, despite of having system-environment interaction producing dephasing, we have shown how to introduce dissipative elements to the open system dynamics due to the interference effects. 

Note that earlier research has focused, e.g., on how open systems lose their quantum properties due to decoherence~\cite{petru} or how the state of an open system changes when using convex combination of dynamical maps~\cite{sau,vina}. Our approach contains a fundamentally new element in this context. Coherent mixing within an interferometric setup allows to display the effects that the interference has on the evolution of open systems. This leads to rich open system dynamics in terms of non-Markovian memory effects and opens new aspects considering the origin of dephasing and dissipation---even when using one of the most basic interferometric setups and a single photon system only. In the future, it will be interesting to include many body aspects into this framework combined with multiport interferometers.

Our results, therefore, open so far unexplored avenues for the control and engineering of open system dynamics. This includes non-Markovian memory effects, where their source originates  from first superposing two paths having different earlier dynamics and then continuing with the system-environment interaction. In general, we hope that our results stimulate further work for understanding rich dynamical features of open quantum systems, how to engineer them, and how to explore fundamental aspects of quantum mechanics by combining the concepts of open quantum systems---beyond their traditional use---with other physical frameworks. For example, here we only introduced the idea of collision models mixing discrete-time and continuous-time dynamics. It would be very interesting to see what kind of open system dynamics longer chains of beam splitters with continuous-time dynamics in-between produce.

\acknowledgements
This work was financially supported by the Magnus Ehrnrooth Foundation
and the Academy of Finland via the Centre of Excellence program (Project no. 312058).


\end{document}